\documentclass[preprint,aps,draft]{revtex4}

\usepackage{graphicx}
\usepackage{dcolumn}
\usepackage{bm}
\usepackage{natbib}
\usepackage[latin1]{inputenc}

\newcommand{\ket}[1]{\left| #1 \right>}
\newcommand{\bra}[1]{\left< #1 \right|}

\begin{document}
\title{Entanglement, chaos and atomic Wigner function of the Dicke model.}
\author{L. Sanz}
\affiliation{Departamento de Física, UFSCar, Caixa Postal 676,
13565-905, S\~ao Carlos, SP, Brasil}
\author{K. Furuya}
\affiliation{Instituto de Física `Gleb Wataghin', Universidade
Estadual de Campinas UNICAMP, Caixa Postal 6165, 13083-970,
Campinas, SP, Brasil}
\begin{abstract}
We calculate the atomic (spin) Wigner function for the single mode
Dicke model in the regime of large number of two-level atoms. The
dynamics of this quasi-probability function on the Bloch sphere
allows us to visualize the consequences of the entanglement process
between the boson and the spin subsystems. Such investigation shows
a distinct localization behavior of the spin state with respect to
the polar and azimuthal Bloch sphere angles. A complete {\em
breakdown of reflection symmetry} in the azimuthal angle is shown in
the non-integrable case, even at short evolution times. Also, in the
classically chaotic situation, the appearance of {\em sub-planck
structures} in the Wigner function is shown, and its evolution
analyzed.

\end{abstract}
\maketitle

\section{Introduction}
The Wigner function (WF)\cite{Wigner32} has been an important tool
in physics, particularly to explore the interface between
classical and quantum world \cite{berry83}. Furthermore, it is
an useful ingredient in the theory of partial loss of
coherence \cite{decoherence}. Recently, advances in the state
reconstruction problems \cite{Welsch} renewed interest on the
WF, particularly after the measurement of negative
quasi-probabilities \cite{leibfried96,bertet02} which are
generally considered as signatures of non-classical states
\cite{leibfried98}.
Also, quasi-probability functions are relevant for the study of
entanglement process of continuous variables, as in the beam
splitters \cite{kim2002}, in order to analyze separability of two-mode
Gaussian states \cite{duan2000,simon2000,marian2001}, and in studies
of teleportation of non-classical states like in Ref.~\cite{lee2000}.
Since the proposal of manifestation of chaos in the process of
decoherence by Zurek and Paz \cite{Zurek95}, the WF has also been
an important tool for studying this issue~\cite{Habib98,Zurek01}.

Here, our focus is the use of the atomic Wigner function to study
the ideal atom-field entanglement process.
Recent experimental progress on arrays of quantum dots and Josephson
junctions has raised the possibility of super-radiance, a
phenomenon already known to arise in a system with $N$ {\sl two-level}
atoms coupled to a field mode described by the Dicke model (DM)
\cite{dicke}. Some of the solid-state systems relevant to quantum
information that can be mapped into DM are:  `phonon cavity quantum
dynamics' \cite{weig04,vorrath04}; those with Josephson Junctions
and quantum dots, with the possibility of multiqubit entanglement
Dicke-like model \cite{lee04}; and the proposal of `circuit
QED' \cite{Blais04}, which has been constructed recently for one qubit
Jaynes-Cummings case \cite{Wallraff04}. Finally, the atom-molecule
coexistence model near a Feshbach resonance of cold fermions in both
strong \cite{barankov} and weak \cite{andreev} coupling has been
proposed as a DM.

We are interested in exploring the atomic (spin) Wigner function dynamics
of the generalized $N$-Jaynes-Cummings model ($N$-JCM) \cite{JCM}
where both Jaynes-Cummings (JC) and anti-Jaynes-Cummings (AJC)
interaction can be present. These interactions between spin and boson
systems has been shown to appear naturally in trapped ions
\cite{CiracAAMOP96,Solano01}, or in cavity QED by means of strong
classical driving field \cite{Solano03}.
Classically the model becomes chaotic
\cite{Milonni83,Lewenkopf91,Aguiar92}, and some of its
manifestation in the entanglement process has been shown in previous
publications \cite{Furuya98,Angelo99,Angelo01,Emary03,Hou04}. To our
knowledge, it is the first time the dynamical evolution of the
$N$-qubits atomic Wigner function is followed in detail during the
entanglement process since it has been introduced
\cite{Arecchi72,Agarwal81,Dowling94}.

Remarkable features in the AWF can be seen in such wave-packet
dynamics, showing dynamical differences between localization of
the angular variables $\phi,\theta$ on Bloch Sphere.
We also explore the dynamics of the negative-valued
parts of the AWF, usually considered as a hallmark of
non-classicality and interference effects. Another notable aspect
is the distinct time evolution of the AWF,
depending on the initial positions chosen for the centers of the
wave packets. Such sensitivity to initial conditions have been
already noted in references \cite{Furuya98,Angelo99}, at the level of
the integrated quantities like the subsystem linear entropy or mean
values, but the details of the behavior of quasi-probability functions
was lacking.

This paper is organized as follows: in Sec.~\ref{sec:theory} we
present the classical analog of the Dicke model to define and
show some quantities used in the following sections. Also, we
review the notion of sensitivity of the entanglement process
on the initial position of the coherent packet, in both
{\em integrable} and {\em chaotic} cases, already found in previous
works. Sec.~\ref{sec:wigners} is reserved to present our results
for the $N$-qubit atomic Wigner function for some selected initial
conditions and analyze its time evolution under the point of
view of the entanglement process. In Sec.~\ref{sec:summary} we
summarize our results.

\section{The model and entanglement dynamics within the large-$N$
approximation}
\label{sec:theory}
We consider a generalized version of $N$-atom (qubit) Dicke model,
with variable coefficient for the rotating and counter-rotating
wave term. Moreover, since our aim here is to pursue the effect of
the classically chaos generating term on the entanglement between
the $N$-atom and the field, we will ignore the interaction between the
atoms (qubits) and treat the atomic system as a large spin ($N=2J$).
\begin{eqnarray}
\hat{H}&=&\hbar\omega_0\hat{a}^{\dagger}\hat{a}+\hbar\omega_a
\hat{J}_z
+ \frac{G}{\sqrt{2J}}\left(\hat{a}\hat{J}_{+}+\hat{a}^{\dagger}
\hat{J}_{-} \right) +\nonumber\\
&& +\frac{G^{\prime}}{\sqrt{2J}}
\left(\hat{a}^{\dagger}\hat{J}_++\hat{a}\hat{J}_-\right).
\label{eq:Hmaserq}
\end{eqnarray}
Here, $\omega_0$ and $\omega_a$ are frequencies associated with free
Hamiltonians for field and atoms respectively. $G$, $G^{\prime}$ are
coupling constants associated with atom-field interaction within the
dipole approximation. The usual Rotating Wave Approximation (RWA) is
recovered by setting $G'=0$. The field observable are described by
means of the creation and annihilation operators $\hat{a}$ and
$\hat{a}^{\dagger}$, whereas $\hat{J}_z$, $\hat{J}_{\pm}$ are
pseudo-spin operators associated to an atomic observable. This model
is used to describe both, cavity QED experiments
\cite{David93,Solano03} (with $G=G^{\prime}$, but usually to an
excellent approximation one can set $G^{\prime}=0$) and trapped-ion
systems. In the last system, interactions with different couplings
$G\ne G^{\prime}$ can be generated \cite{CiracAAMOP96,Solano01}.

For the purpose of our study where both systems are initially
pure and separable in quasi-classical states, the appropriate initial
state $\ket{w \nu}$ is a product of the field and
atomic coherent states defined
as~\cite{Glauber63,Arecchi72,Zhang90}:
\begin{eqnarray}
\ket{\nu}&=&\hat{D}\left(\nu\right)\ket{0}=e^{-\frac{|\nu|^2}{2}}
e^{\nu\hat{a}^{\dagger}+\nu^{*}\hat{a}}\ket{0} \nonumber\\
\ket{w}&=&\left(1+|w|^2\right)^{-J}e^{w\hat{J}_+}\ket{J,-J}.
\label{eq:DEFec}
\end{eqnarray}
Here, $J=N/2$ and the variables $w$ and $\nu$ can be written as
a function of the classical variables in the corresponding phase
spaces, $(q_f,p_f)$ for the field, and $(q_a,p_a)$ for the atomic
degree of freedom
\begin{eqnarray}
w&=&\frac{p_a+\imath q_a}{\sqrt{4J-\left(p^2_a+q^2_a\right)}}
\nonumber\\
\nu&=&\frac{1}{\sqrt{2}}\left(p_f+\imath q_f\right)
\label{eq:PARA}
\end{eqnarray}
A corresponding classical Hamiltonian can been obtained by a
standard procedure~\cite{Saraceno}, using the above defined
coherent states $\bra{w\nu}\hat{H}\ket{w\nu}$~\cite{Aguiar92}
\begin{eqnarray}
{\cal H}\left(q_a,p_a,q_f,p_f\right)=\frac{\omega_0}{2}\left(
p_f^2+ q_f^2\right)+\frac{\omega_a}{2} \left(p_a^2+q_a^2\right)\\
-\omega_aJ+\frac{\sqrt{4J-\left(p_f^2+q_f^2\right)}}{4J}\left(
G_{+} p_ap_f+G_{-}q_aq_f\right),\nonumber
\label{eq:Hmaserc}
\end{eqnarray}
with $G_{\pm}=G\pm G^{\prime}$. The classical dynamics associated
with this Hamiltonian were explored before \cite{Aguiar92}, and
shown that:{\sl (i)} integrable situations are recovered when either
$G$ or $G^{\prime}$ is zero; {\sl (ii)} the most chaotic dynamics is
associated with the condition $G=G^{\prime}\approx
\mathcal{O}(\omega_0, \omega_a)$, as we increase the coupling
constant. Our aim is to investigate the time evolution of the
initially quasi-classical wave-packet along with the occurrence of
entanglement between the $N$-atom and the field systems.
Particularly, we are looking for the possible differences in the
reduced wave-packet dynamics when we compare {\sl integrable} and
{\sl chaos generating} interactions. The connection with the
classical dynamics is established by choosing coherent states as
initial states, centered at the corresponding points of the phase
space. Then, we let the system evolve by means of the Hamiltonian
(\ref{eq:Hmaserq}) and explore the entanglement dynamics solving
numerically the Schrödinger equation.

\begin{figure}[h]
\vspace{-0.5cm}
\includegraphics[scale=0.2]{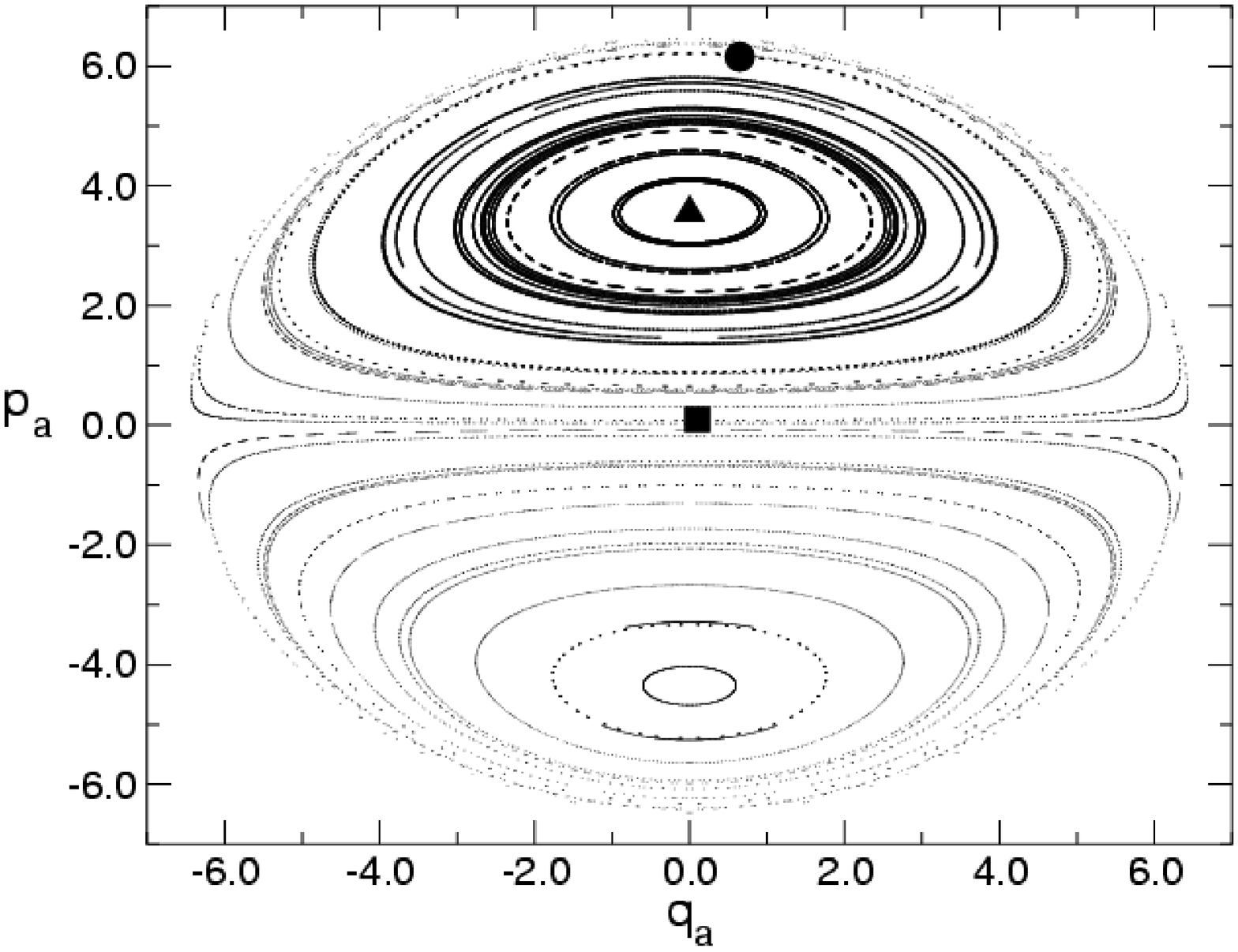}
\includegraphics[scale=0.24]{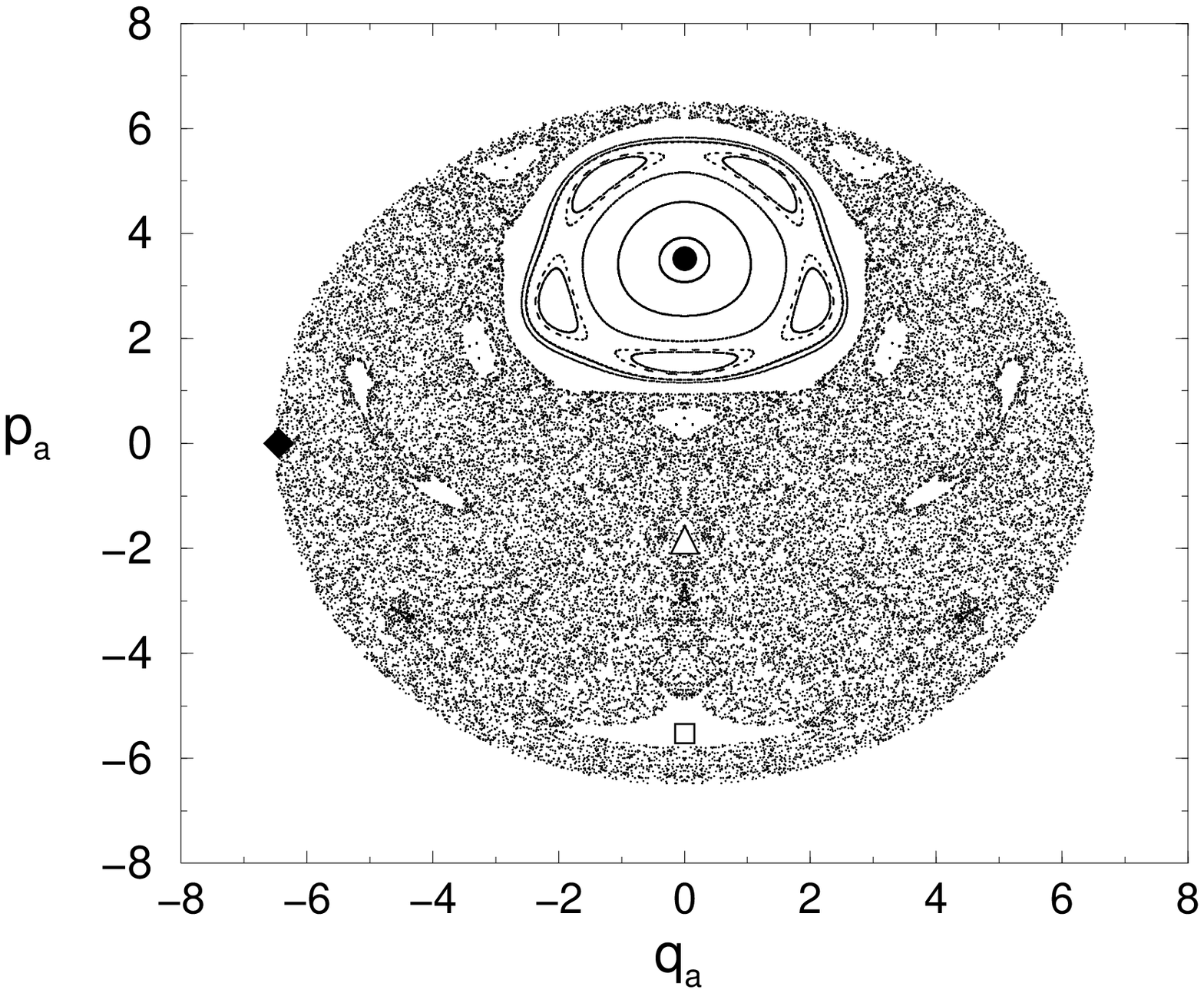}
\caption{Poincaré section for the atomic degree of freedom (with
$q_f=0.0$ and $p_f>0$) in the resonant case $\omega_a=\omega_0$
and energy $E=21.$, with $N=2J=21$. Left: integrable case with
$G=0.5$ and $G^{\prime}=0.0$. Right: non-integrable case with
$G=0.5$ and $G^{\prime}=0.2$.}
\label{fig:poincare}
\vspace{-0.5cm}
\end{figure}

In order to know where to put the initial atomic and field
wave-packets, the first step is study the structure of classical
phase space. This can be done by using the Poincaré section. In
Fig.~\ref{fig:poincare}, we plot the projection of Poincaré
section in the atomic plane ($q_a$,$p_a$) for $q_f=0.$ and
$p_f>0$. The surfaces of section correspond to both,
the right one to the integrable case with ($G=0.5$,
 $G^{\prime}=0$) and the left one to the soft chaos ($G=0.5$,
$G^{\prime}=0.2$). Here, and along the present work, the total
energy is fixed at $E=2J=21.0$ with $J=10.5$.
The coupling values corresponds to the non-super-radiant phase
($G_+ < 1$)~\cite{Aguiar91QO}. The limit of atomic phase space
is indicated by a border at radius equal to $\sqrt{4J}$.
Integrable section shows a separatrix of motion along the line
$p_a=0.0$ and concentric tori around each of the two stable
periodic orbits. A pro-eminent feature of the non-integrable
surface of section is a large stability island for $p_a>0$.

The symbols in Fig.~\ref{fig:poincare} show the chosen centers
for the atomic coherent states. We choose two specific initial
conditions (i.c.) for each case. In the integrable case, the
first one (I1) is on an internal tori belonging to the region
$p_a>0$, marked by a triangle; and the second (I2), marked by
a circle, located near the border of the atomic phase space.
For soft chaos situation, the first i.c. (N1) where chosen on
a point inside the largest stability island (circle).
The second condition (N2) is located in the chaotic sea
(triangle). Specific values of $q_a$, $p_a$ and the mean value
of $\hat{J}_z$ operator for the associated atomic coherent
state are listed in Table~\ref{tab:condition}.
\begin{table}[h]
\begin{center}
\begin{tabular}{|l|c|c|c|}
        \hline
I.C. & $q_{a}/\sqrt{4J}$ & $p_{a}/\sqrt{4J}$&$\langle\hat{J}_z
\rangle/J$\\
         \hline \hline
I1&$0.0$&$0.55$&-0.43\\
        \hline
I2&$0.1$&$0.95$&0.83\\
        \hline
N1&$0.0$&$0.54$&-0.47\\
\hline
N2&$0.0$&$-0.28$&-0.84\\
\hline
\end{tabular}
\caption{$q_a$, $p_a$ and $\langle\hat{J}_z\rangle$ values for the chosen
initial conditions.}
\end{center}
\label{tab:condition}
\end{table}
%

\begin{figure}[h]
\includegraphics[scale=0.5]{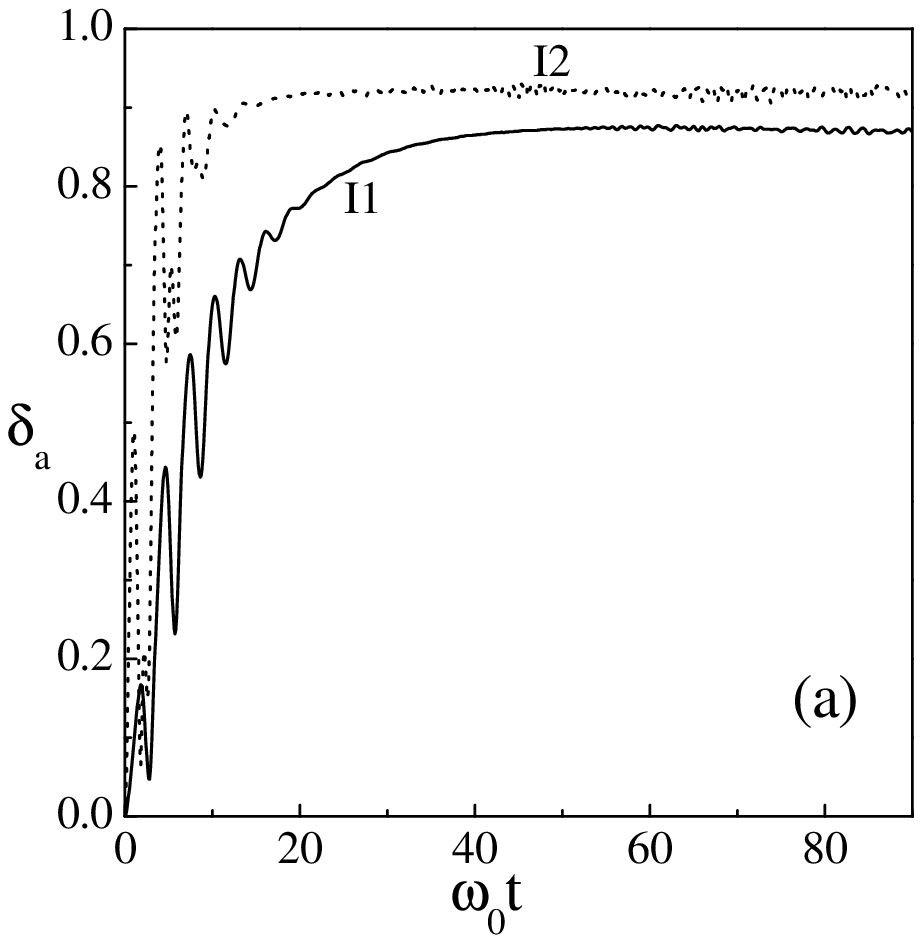}
\includegraphics[scale=0.5]{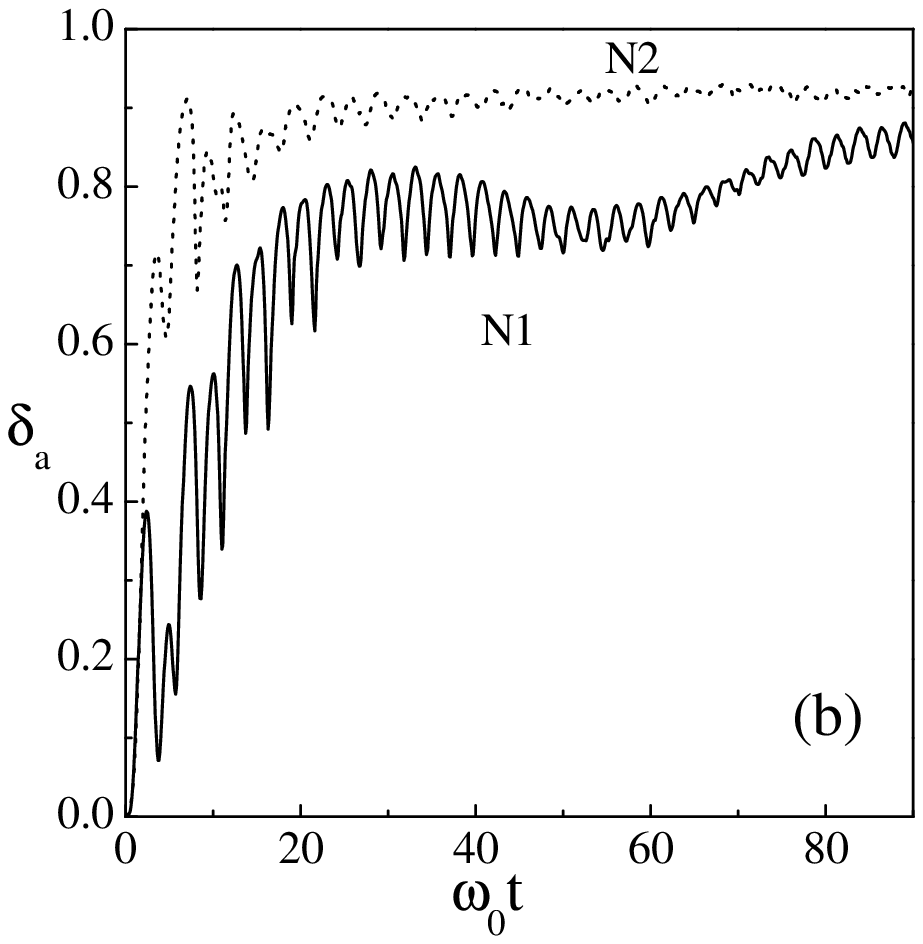}
\caption{Atomic Linear Entropy of the $N$-JCM associated with
the four initial conditions centered at the positions shown in
Fig.~\ref{fig:poincare} (with the same parameter values) and
listed in Table~\ref{tab:condition}.
(a) Integrable case: i.c. I1 (Solid line) and I2 (dotted line);
(b) Soft chaos: i.c N1 (Solid line) and N2 (dotted line).}
\label{fig:entropies}
\end{figure}

As a second step, one should calculate the time evolution of the
atomic linear entropy (ALE) which, in this globally pure
bipartite system, can be used as a measure of entanglement.
This calculation is done in three stages:
First we diagonalize the Hamiltonian~(\ref{eq:Hmaserq})
numerically and use the eigenvalues and eigenstates in order to
find the temporal total density operator, $\hat{\rho}\left
(t\right)$. Then, we calculate the reduced density operator of
the atomic subsystem,
$\hat{\rho}_a=\mbox{Tr}_f\left[\hat{\rho}\left(t\right)\right]$,
and finally  obtain $\delta_a\left(t\right)= 1-\mbox{Tr}_a
\left[\hat{\rho_a}^{2}\left(t\right)\right]$. Evolution in time
of the ALE for i.c.'s I1 and I2 are plotted in
Figure~\ref{fig:entropies}(a). For both i.c.'s, the subsystem
entropy increases in the mean as time goes on, until a plateau
is reached. The details of the curve such as the specific values
 of the ALE in the plateau, the particular oscillatory behavior
and the entanglement rate depend on each initial condition. In
particular, we observe that those atomic i.c.'s with the
classical dynamics restricted to a well delimited region in
phase space on the tori region are more resistant to entangle
with field. This relation between fast entanglement process and
less localized classical dynamics was pointed out in a previous
work~\cite{Angelo99}.

The ALE for the soft chaos situation are plotted in
Fig.~\ref{fig:entropies}(b). In this plot, it is clear that
larger entanglement rate is also associated to the chaotic
i.c.'s. It is interesting to compare our results for I1
and N1, solid lines in Fig.~\ref{fig:entropies}(a,b). For I1,
ALE shows more regularity in the oscillations and reaches the
corresponding plateau around $t \approx 30$,
whereas in ALE for N1 the oscillations are less regular and
takes a longer time to reach the plateau, at $t \approx 70$. Also,
ALE for N1 condition keeps a small oscillation around its mean value (due to the
non-RWA term) and one can also see a certain large period
modulation in contrast to I1. The relation between the behavior
of the classical trajectory and maxima and minima of $\delta_a$
was studied in Ref.~\cite{Angelo01}, and we will not discuss
here.
Instead, the atomic Wigner Function (AWF) is used to visualize the
behavior of global state as projected in atomic phase space.
In the next section, we will illustrate the following aspects:
{\sl (i)} that the AWF provide us the information that those
i.c.'s that are more ``protected'' against entanglement process,
have a strong localization on the Bloch sphere; {\sl(ii)} we
also show how the destruction of tori due to chaos in the
classical dynamics goes along with the delocalization of the
quantum wave-packet during its temporal evolution.

\section{Dynamics of atomic Wigner function.}
\label{sec:wigners}
The phase space quasi-probability distributions of
electromagnetic field and atom have been discussed by several
authors~\cite{Wigner32,Gerry97,Stratonovich57,Arecchi72,
Agarwal81}. Here, we adopt the definition of Wigner function
in terms of arbitrary angular momentum basis, as introduced by
Agarwal~\cite{Agarwal81}. This function is defined as
\begin{eqnarray}
W\left(\theta,\phi,t\right)=\sqrt{\frac{2J+1}{4\pi}}\sum
\limits_{K=0}^{2J} \sum\limits_{Q=-K}^{K}\varrho_{KQ}\left(t
\right)Y_{KQ}\left(\theta,\phi \right),
\label{eq:FWatomica}
\end{eqnarray}
where $\varrho_{K,Q}$ is given by
\begin{eqnarray}
\varrho_{K,Q}\left(t\right)=\mbox{Tr}\left[\rho_a\left(t\right)
\hat{T}_{KQ}\right].
\label{eq:mcaractA}
\end{eqnarray}
This is the characteristic function associated with atomic
density operator $\rho_a\left(t\right)$. Here, $\hat{T}_{KQ}$
is the multipole operator acting in the angular momentum
space~\cite{AngularM}
\begin{eqnarray}
\hat{T}_{KQ}&=&\sum\limits_{M=-J}^{J}\left(-1\right)^{J-M}
\sqrt{2K+1} \left(\begin{array}{ccc}
J&K&J\\
-M&Q&M-Q
\end{array}
\right) \nonumber\\
&&\times\ket{J,M}\bra{J,M-Q}.
\label{eq:Tkq}
\end{eqnarray}
In Eq.(\ref{eq:FWatomica}), the usual Wigner $3J$ symbol
 has been used, and $Y_{KQ}\left(\theta,\phi\right)$ indicates
the spherical harmonics defined over the Bloch sphere. There,
$\theta$ is the polar angle and $\phi$ is the azimuthal angle.
The distribution of any angular momentum state can be studied
using the AWF. As shown in Dowling {\it et al.}~\cite{Dowling94},
 it is possible to estimate the indeterminacy in the measure of
the atomic observable $\hat{J}_z$, $\hat{J}_x$ $\hat{J}_y$  in the
state through its variances . In fact, if the state has a large
probability associated to a well-defined eigenvalue of $\hat{J}_z$,
its atomic Wigner function shows a strong localization in polar
angle $\theta$. In a similar way, indeterminacy associated to the
measure of $\hat{J}_x$ and $\hat{J}_y$ means ignorance on the
azimuthal angle $\phi$.

In order to obtain the AWF, we use our previous results of atomic
density matrix operator. Because the basis used was the Dicke
states $|\hat{J},\hat{J}_z\rangle$, we can calculate the action
of $\hat{T}_{KQ}$ on each atomic density matrix elements,
obtaining $W\left(\theta,\phi,t\right)$. Also, we always set
the $\phi=0$ value exactly at the center of each initial atomic
coherent packet. That means, if the wave packet is not on the
$X$-axis, we rotate the $XY$ plane by a certain angle
$\phi(t=0)=\phi_0$ in such a way that the direction defined by
the vector $(\sin\phi_0,-\cos\phi_0,0)$ coincides with the
rotated $X$-axis.
\subsection{ Integrable case with $G=0.5$ and $G^{\prime}=0$.}
\label{subsec:integrable}

\begin{figure}[h]
\includegraphics[scale=0.5]{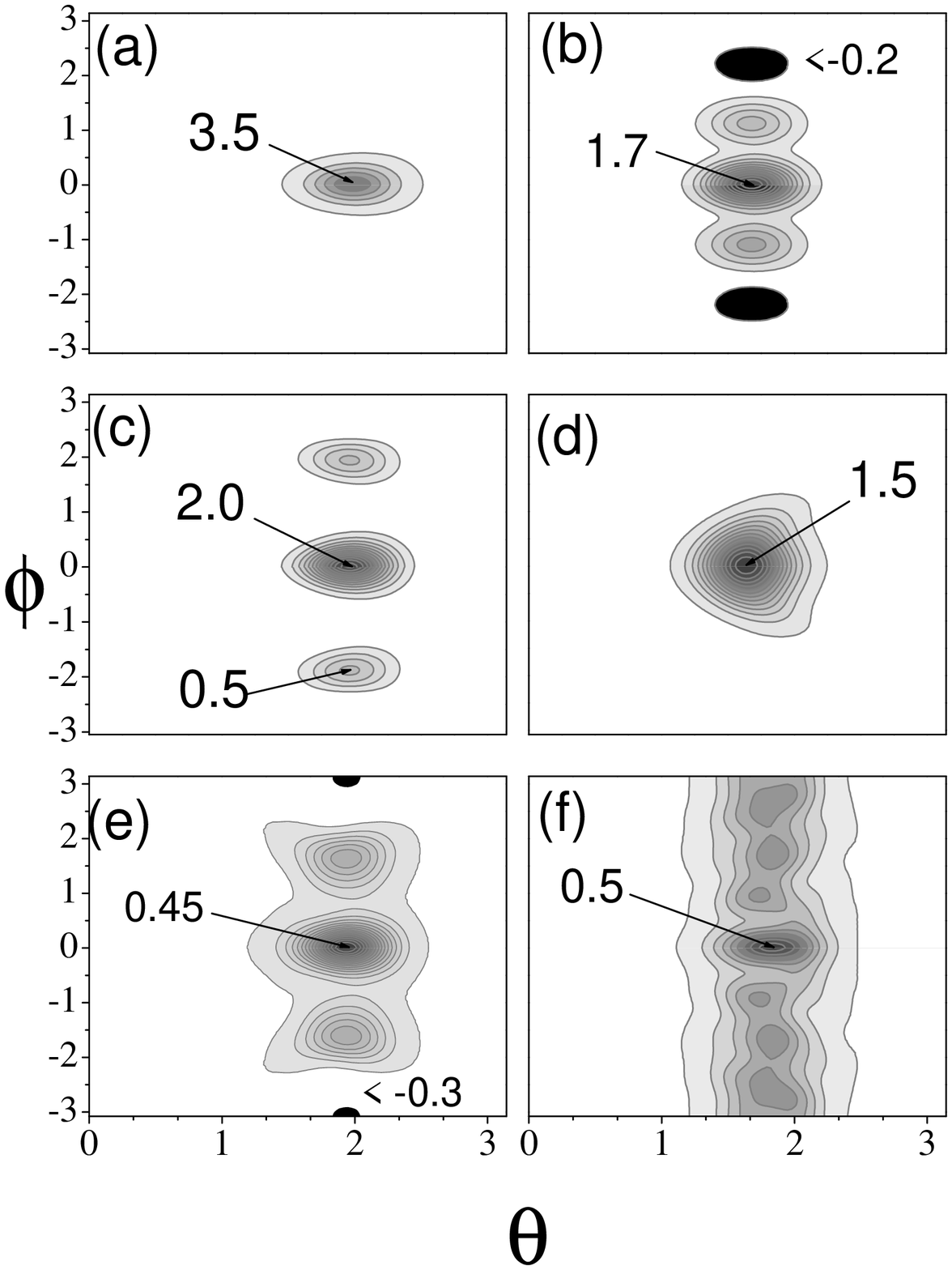}
\caption{Shaded contour plots of atomic Wigner function
associated with the i.c. I1: (a) initial time; (b) {\em first maxima}
of the corresponding atomic linear entropy (ALE); (c){\em first
minimum} of ALE; (d) {\em second maxima}; (e) {\em second minima};
(f)$t=40$ at the {\em plateau region}. Negative valued regions of
AWF are drawn in black.}
\label{fig:wai1}
\end{figure}

For the integrable case, we show the
snapshots of the temporal evolution of contour lines of AWF in
Fig.~\ref{fig:wai1} for the i.c. I1. The initial coherent state,
Fig.~\ref{fig:wai1}(a), has its maximum value ($3.5$) at
$\theta=0.64\pi$, which corresponds to
$\langle\hat{J}_z\rangle\approx-0.43J$. At the time when the
first maxima of ALE is reached, the atomic state has a more
delocalized distribution, shown in Fig.~\ref{fig:wai1}(b), with
two negative valued regions (in black). A formation of three
overlapping positive peaks starts, with maxima (almost) at the
equator of Bloch sphere ($\theta=\pi/2$). At this time, AWF has
a maximum value lower than the one at the initial time ($\approx
 1.7$). The appearance of a negative part in the AWF with value
$\approx -0.2$ ($10\%$ of maximum value) indicates the
non-classical character of this state.

The delocalization of the state in the azimuthal angle is
associated with the increase of the ALE. This assertion can be
confirmed by checking the evolution for consecutive maxima and
minima. The forms of AWF are shown at times corresponding to
the {\em first minimum}, Fig.~\ref{fig:wai1}(c); {\em second maximum},
Fig.~\ref{fig:wai1}(d), and {\em second minimum},
Fig.~\ref{fig:wai1}(e). Comparing them, it is clear that the
AWF is more localized in the $\phi$ variable at times when ALE
has minima. Although in Fig.~\ref{fig:wai1}(d), the three peaks
have coalesced into one, the packet sweeps a larger interval
over $\phi$ values than those found at the first minimum time
scale. We can also observe that AWF maximum value oscillates
around $\theta\approx 1.8$. A negative valued portion is still
present but it became significantly smaller (less than $1\%$)
than in Fig~\ref{fig:wai1}(b), so it is not possible to see in
Fig~\ref{fig:wai1}(c,d), but reappears in Fig.~\ref{fig:wai1}(e).
From this sequence, it is clear that atomic state looses both,
azimuthal and polar localization, associated with the increase in
the atomic linear entropy. However, at those times when the ALE
plateau is reached, we see how the AWF (plotted in dotted lines)
still has localization in $\theta$ (near the equator in the Bloch
sphere), having non-zero values only in the interval $1.6 \le \theta
\le 2.5$. This shows that for the particular initial condition
considered on an internal torus (distant from the separatrix
motion and the border of the phase space), an increase in the
ALE is associated with the increasing delocalization with
respect to the azimuthal angle much more than to the polar
variable.

\begin{figure}[h]
\includegraphics[scale=0.5]{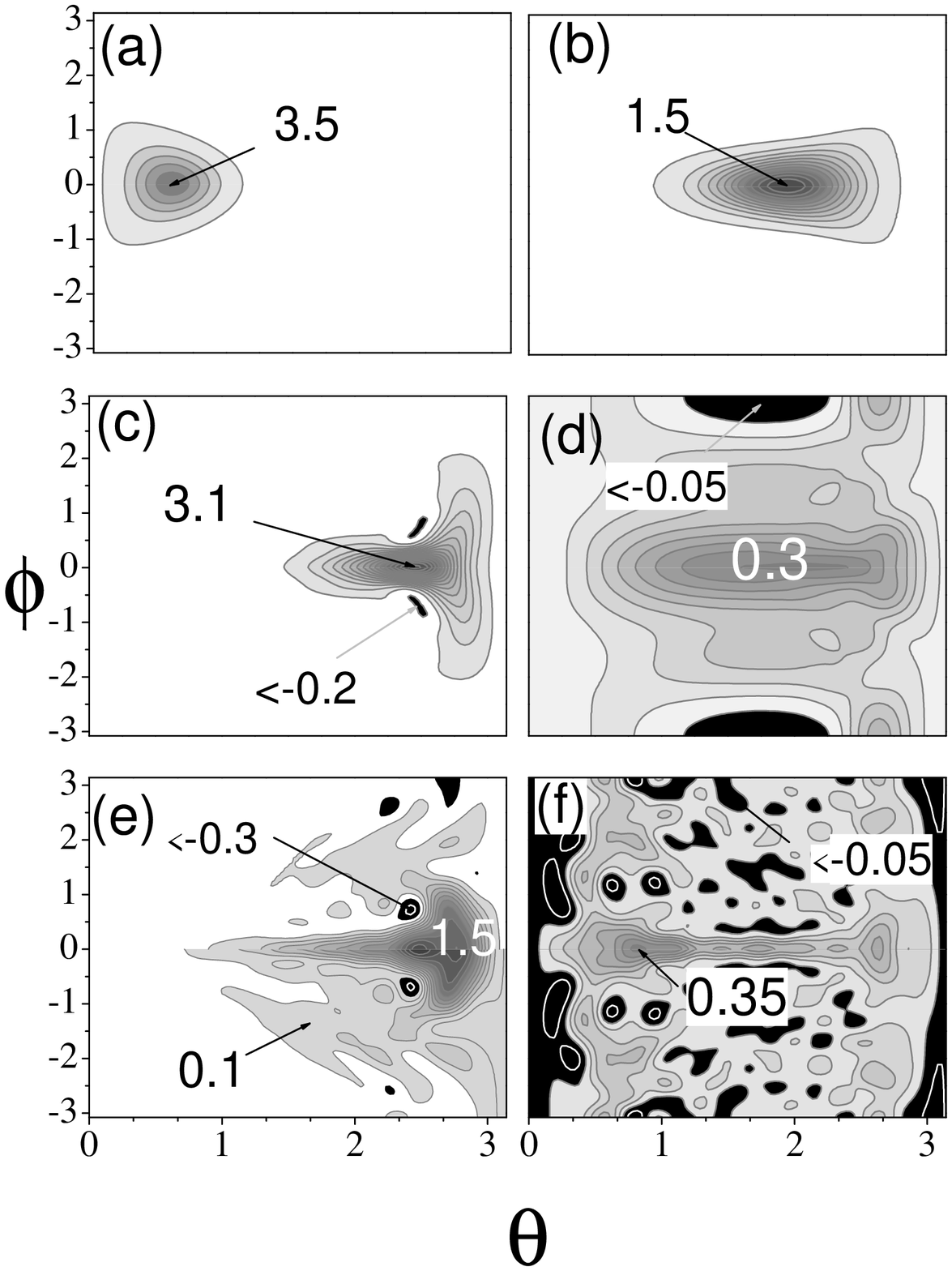}
\caption{Shaded contour plot atomic Wigner function associated
with the i.c. I2: (a) initial time; (b) {\em first maxima} of
corresponding atomic linear entropy (ALE); (c) {\em first minimum}
of ALE; (d) {\em second maxima}; (e) {\em second minima}; (f)$t=40$
at the {\em plateau region}. Negative valued regions of AWF are
drawn in black.}
\label{fig:wai2}
\end{figure}
%
Now we shall compare previous case with the time evolution of the
AWF for condition I2 ($\theta\approx 0$) which is close to the
largest value of $\langle\hat{J}_z\rangle$, shown in
Fig.~\ref{fig:wai2}. At the initial time, Fig.~\ref{fig:wai2}(a),
the AWF is well localized in both angular variables and, as the
two subsystems interact, we can see that the AWF begins to lose
localization mostly in the polar variable (non-zero values in the
interval $1.0 \le \theta \le 2.7$), at the time scale of the
first maximum in ALE. Physically, the spreading of the AWF and
the increase in the atomic linear entropy, tell us that for this
 i.c. (I2), the increasing entanglement of the $N$-atoms with
the field is mostly associated with an increase of the
participating states of atomic levels in the $\hat{J}_z$
spectrum (shown by the spreading of the AWF in $\theta$
variable). The value of the ALE at the first maximum for I2
($\delta_a \approx 0.5$) is greater than for I1 ($\delta_a$
lower than $0.2$), where the increasing delocalization is
primarily in the azimuthal variable $\phi$.

At the time corresponding to the {\em first minimum} of the ALE,
Fig.~\ref{fig:wai2}(c), we observe that AWF forms a three-peak
structure seen in the I1 case. However, there is
a appreciable spreading in the $\phi$ variable but not allowing
the peaks to become separated, and near the central position in
$\phi=0$ occurs also a notable delocalization in $\theta$
variable. The negative part is no longer localized as in the
previous i.c., and its value is $\approx 10\%$ of the maximum of
AWF. Again, we observe the connection between minima of
$\delta_a$ and the angular localization: the AWF for the second
maximum of ALE, Fig.~\ref{fig:wai2}(d), shows how the atomic
state is completely delocalized in both variables at this time.
The AWF is non-zero practically at all points on the atomic
phase space. This situation is reversed at the time of the
second minimum of ALE, with a tentative to re-gain some
localization in $\theta$ variable and a structure which roughly
resembles Fig.~\ref{fig:wai2}(b). It is also interesting to see
how the negative part of AWF reappears and it is even more
pronounced than in the previously referred time ($\approx
20\%$). At times when the plateau is reached the AWF is totally
delocalized and not even a signal of a main positive peak is
present, which was the case we found in the plateau times of
the internal torus case. Negative part is less than $1\%$ of
maximum value of AWF.

From this results, it is clear that initial condition I1
dynamically preserves the localization of the AWF, especially in
the polar angle. This is associated with a certain {\em inhibition
in the entanglement process}. The difference on the value of the
atomic linear entropy between the two initial conditions presented
here is clearly related with the delocalization process in both
azimuthal and polar angles. Hence, the dynamics of the internal
tori is protected against the entropy increase, and this is
related with the localization in the polar angle; whereas, the
i.c.'s located near the separatix and the border do not have
such a dynamical protection. Other initial coherent states with
similar characteristics has qualitatively analogous behavior
for the ALE.

Another characteristic is the clear appearance of some
{\sl sub-planck structures}, namely the structures with considerable
amplitudes with their supports in areas much smaller than
$\hbar$ in phase space, similar to those discussed by
Zurek~\cite{Zurek01}. On the Bloch sphere, the minimum action
area ($\hbar$ is taken to be $1$ here) is associated with the
size of atomic coherent state (at $t=0$) which defines a
minimum-uncertainty packet.
 This minimum action area can be inferred, for instance, in
Fig.~\ref{fig:wai1}(a) and Fig.~\ref{fig:wai2}(a) for N=21.
Such sub-planck structures which are peaks confined in areas
significantly smaller than the size of the initial packet in
the ($\phi,\theta$)-plane appear for instance in Figs
\ref{fig:wai2}(f), \ref{fig:wan1}(d,f) and \ref{fig:wan2}(f).
It is interesting to note that its appearance is indeed
connected with the time where the atomic subsystem has lost
its own coherence by entangling with the field. What is
remarkable is that, for the integrable case, the entanglement
process leaves the AWF with sub-planck structure {\sl only} for
i.c. I2 but not for I1. This is, to one side, AWF counterpart
of the {\sl rapid loss of coherence} that occurs for the wave packet
located near the separatrix of motion \cite{Angelo99}, but it
is more than simply accelerating the entanglement process:
the dynamical instability also generates structures similar to
chaotic case as we shall see in the next subsection.
\subsection{Non-integrable case: $G=0.5$ and $G^{\prime}=0.2$}
\label{subsec:non-int}

\begin{figure}[h]
\includegraphics[scale=0.5]{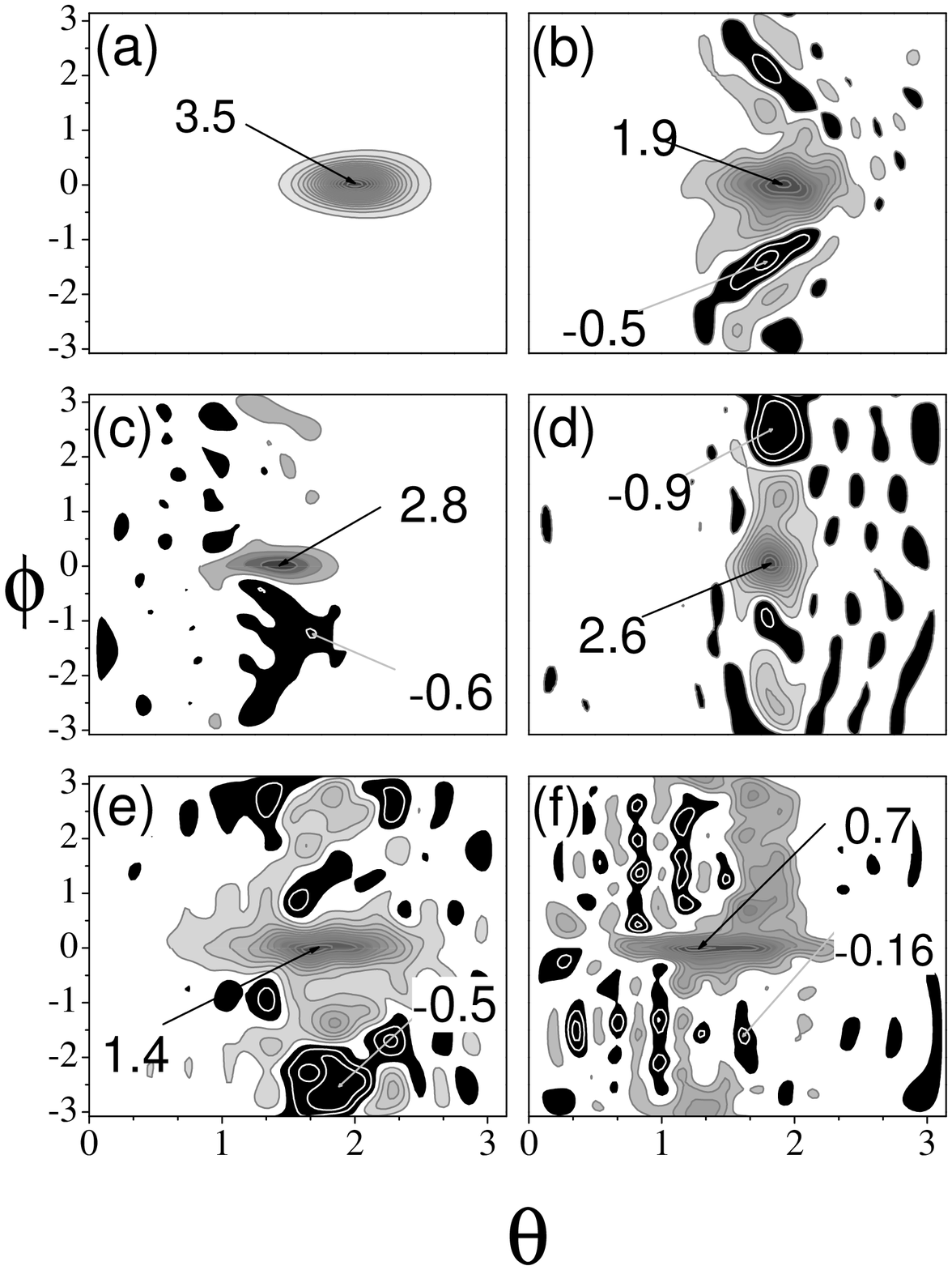}
\caption{ Shaded contour plot of the atomic Wigner function
associated with i.c. N1: (a) initial time; (b) {\em first maxima} of
corresponding atomic linear entropy (ALE); (c) {\em first minimum}
of ALE; (d) {\em second minimum}; (e) {\em third maxima}; (f)
$t=70$ at the {\em plateau region}. Negative valued regions of AWF
are drawn in black.} \label{fig:wan1}
\end{figure}

\begin{figure}[h]
\includegraphics[scale=0.5]{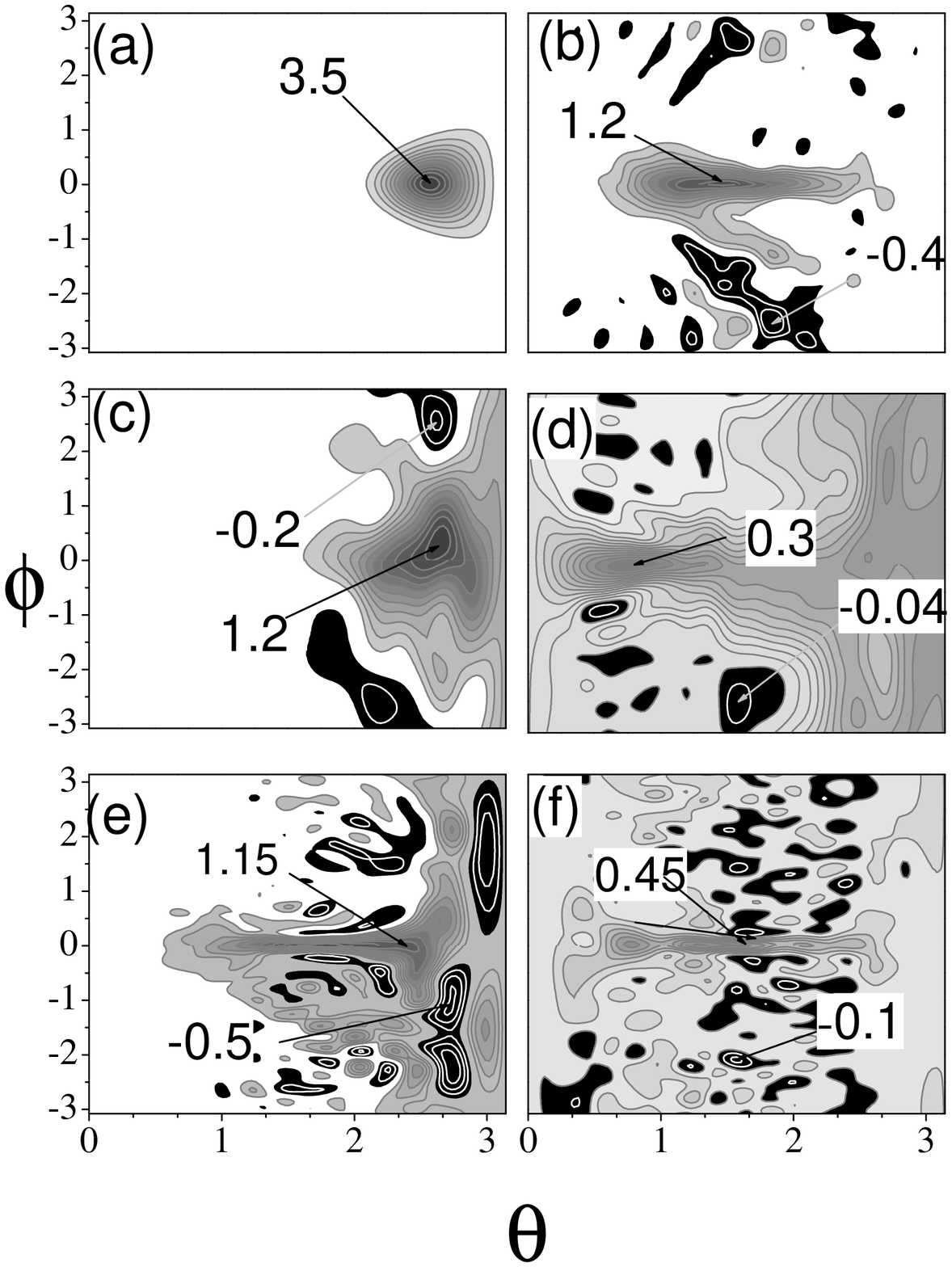}
\caption{ Shaded contour plot of the atomic Wigner function
associated with i.c. N2: (a) initial time; (b) {\em first
maxima } of the corresponding atomic linear entropy (ALE);
(c) {\em first minimum} of the ALE; (d) {\em second maxima}; (e)
{\em second minima}; (f) $t=70$ at the {\em plateau region}. Negative
valued regions of AWF are drawn in black.} \label{fig:wan2}
\end{figure}

Now, we present in Figures~\ref{fig:wan1}-\ref{fig:wan2} our
results for the time evolution of the AWF corresponding to the
{\sl non-integrable} case for the conditions N1 (regular region)
and N2 (chaotic region). Some similarities between integrable and
non-integrable cases can be noticed: first, the connection
between oscillatory behavior and a delocalization-localization
of AWF are still present even in the chaotic i.c. (N2). This
can be seen, for example, in Figs.~\ref{fig:wan2}(c, e), as
compared with Fig.~\ref{fig:wan2}(d). There, AWF seems to
suffer a ``recoil" to a restricted area in ($\phi,\theta$)
plane at times which correspond to a minimum in the ALE
(corresponding to the behavior of Figs.~\ref{fig:wai2}(c, e)
compared with Fig.~\ref{fig:wai2}(d) of the integrable case).
Second feature is related with the similar
``tori protection'' that was found in the integrable case.
Comparing the forms of AWF in the plateau region,
Fig.~\ref{fig:wan1}(f) and Fig.~\ref{fig:wan2}(f), a more
localized AWF (particularly for the positive-valued part) is
evident for the first condition (N1) inside the large stability
island than the second one (N2) in the chaotic region. Notice
that, we also obtain a certain difference for the ALE plateau
values in Fig.~\ref{fig:entropies}(b).

The most interesting aspects are the dynamical differences between
the two cases. Notice that, in the {\sl integrable} case, AWF has an
azimuthal mirror symmetry: $\phi \rightarrow -\phi$, the $\phi< 0$
region being a mirror image of the $\phi> 0$ region. This symmetry
is not present in the {\sl non-integrable} case. At this point, it
is important to recall that this symmetry breaking is already
observed in the classical Poincaré section in atomic phase space
(see Fig. \ref{fig:poincare}). Since we are studying a situation
within the large-$N$ limit, this symmetry breaking can be associated
with {\sl quantum chaos} at the level of spectral distribution
\cite{Lewenkopf91}. Another distinguishable feature is the behavior
of the sub-planck structures in the AWF. They have appeared in
Fig.~\ref{fig:wan1}(b), in spite of the tori protection, and remain
for times at the plateau region. The size of this sub-planck
structures seems to saturate after the entanglement time, confirming
for the present model the results shown by Zurek. Also, it is
notable the presence of a larger number of negative sub-planck
packets than positive ones, although we do not have any explanation
for this fact.
\section{Summary}
\label{sec:summary}

This work gives a complete analysis of the temporal behavior of the
entanglement process in the $N$-JCM in the large-$N$ wave packet
dynamics. Previous results have pointed out the sensitivity to
initial conditions of the atomic linear entropy. Here, a calculation
of the atomic Wigner function allowed us to uncover additional
information about the atomic subsystem, not visible in an integrated
quantities like the entropy. This allowed us to have a better idea
of what is happening to the reduced atomic state during the
entanglement process, as a function of both, the type of
interactions present (rotating and counter-rotating) and the initial
position of the coherent wave-packet.

A very conspicuous information obtained in this way, is the dynamics
of the amount of the delocalization of the AWF during the
entanglement process as a function of both angular variables on the
Bloch sphere. Also, we show that the presence of the classical tori
structure in the phase space surrounding the center of the coherent
wave packet, is an indication at the quantum level of a certain
inhibition in the coherence loss. Thus, by breaking the
integrability we also break this protection against delocalization
in the polar angle for the initial coherent state centered on the
internal tori. However, the regular surrounding is still an
indication of slower loss of coherence and, apparently the larger
the island of stability around the wave packet, stronger is this
effect on the quantum wave packet. The symmetry breaking of the AWF
for any time ($t>0$) is another characteristic of the non-integrable
case.

The most interesting aspect is the dynamics of the sub-planck
structures: it is completely absent in the regular initial
conditions of the integrable situation, but do appear in the
long-time (ALE plateau region) behavior for the packet placed
near the separatrix of motion. In the non-integrable case, even
the wave packets placed inside the regular island do develop
sub-planck structures well before the plateau of the ALE is
reached, and the chaotic cases show such structures already at
the first maximum of the ALE. Such sub-planck structures seems
to be directly associated with the destruction of the ``tori
protection'' and faster entanglement due to less restricted
dynamics in phase space, thus being an indicator of dynamical
instability connected to {\sl quantum chaos}.
\begin{acknowledgments}
It is a pleasure to acknowledge R. M. Angelo for many helpful
discussions.
We thank financial support from FAPESP (Fundação de Amparo à
pesquisa do Estado de São Paulo) under grant 03/06307-9 and
CNPq (Conselho Nacional de Pesquisa, Brazil) under grants
146010/99-0 and 300651/85-6.
\end{acknowledgments}

\end{document}